# How science maps reveal knowledge transfer: new measurement for a historical case

**Sándor Soós**
Dept. of Science Policy and Scientometrics, Library and Information Centre of the Hungarian Academy of Sciences (MTA)

**Abstract**

Modelling actors of science via science (overlay) maps has recently become a popular practice in Interdisciplinarity Research (IDR). The benefits of this toolkit have also been recognized for other areas of scientometrics, such as the study of science dynamics. In this paper we propose novel methods of measuring knowledge diffusion/integration based on previous applications of the overlay methodology. New indices called *Mean Overlay Distance* and *Overlay Diversity Ratio*, respectively, are being drawn from previous uses of the Stirling index as the main proxy for knowledge diversification. We demonstrate the added value of this proposal via a case study addressing the development of a rather complex discourse in biology, usually referred to as the *Species Problem*. The selected topic is known for a history connecting various research fields and traditions, being, therefore, both an ideal and challenging case for the study of knowledge diffusion.

**Introduction**

Recent developments in the field of science mapping induced a variety of applications within evaluative as well as within structural scientometrics. The so-called *Science Overlay Map* technique or toolkit has been introduced by Rafols, Porter and Leydesdorff (2010), upon a mapping exercise of global science in terms of Web of Science Subject Categories or SCs (Leydesdorff and Rafols 2009). Via this toolkit, any collection of (WoS-indexed) publications can be represented as an overlay on the global map (hence its name), representing, therefore, its field composition and position on the scientific landscape. Consequently, countries, institutions, researchers, research topics or any other meaningful aggregations can be profiled and compared through this structural model.

An outstanding feature of the overlay toolkit is its capability to convey rich structural information on research profiles. The global science map (basemap) involved in profile mapping is a proximity network of research fields (SCs), based, in principal, on the bibliographic coupling of Subject Categories. Consequently, overlay maps for any aggregate of papers not only encode for the distribution of the aggregate over current fields of science, but also for the relation (cognitive distance) of the fields included in the overlay map. This nice feature is exploited with an ever increasing interest in applying this model within *Interdisciplinarity Research* (IDR). In IDR special focus is being put on inventing measures that summarize this multifaceted information within overlay maps in order to quantify the multi- and/or interdisciplinarity of research profiles. The most popular overlay-based measure of multidisciplinarity so far is the so-called generalized *Stirling index* (Stirling 2007) to be drawn upon any overlay map. Given a set of papers ranging over *n* Subject Categories, the measure takes the form

$$\text{Stirling index} = \sum_{i=1, j=1}^{n} p_i p_j d_{ij}, \text{ whereby}$$

- $p_i$ is the weight of the *i*-th Subject Category $i = 1, \ldots, n$,
- $p_i$ is the weight of the *j*-th Subject Category $i = 1, \ldots, n$,
- $d_{ij}$ is the distance of the *i*-th and the *j*-th Subject Category as determined by the basemap for the overlay.

The Stirling index can be interpreted as a measure of multidisciplinarity, capturing at least three aspects of cognitive diversity: the variety, the balance and the disparity of fields within research profiles (Leydesdorff and Rafols 2009). The index is rather flexible, as each parameter can be evaluated with different indicators: SCs can be weighted along by their relative frequency within the aggregate, but also with e.g. the impact of the associated papers; similarly, the distance term can be interpreted with a series of network measures, allowing for a variety of aspects to be quantified (Soós and Kampis 2011, 2012). The proposal discussed below highly depends on this IDR methodology: in what follows, by the *overlay toolkit* we mean both the *Science Overlay Map* model and the associated structural measures.

*An overlay-based model of science dynamics*

A huge potential in the approach outlined above is the application of the overlay toolkit in modelling the dynamics of science. In particular, two fundamental processes driving the development of scientific knowledge, namely, (knowledge) *diffusion* and (knowledge) *integration* seem to be outstandingly well characterizable via the overlay methodology. Diffusion and integration, in this context, are being conceptualized as converse processes: diffusion occurs when a subject propagates through a variety of research fields, yielding multi- or interdisciplinary composition of the research topic; integration, at the other extreme, is conceived as research undertaken in various fields converges towards a synthesis, often exhibited by the emergence of a novel field (for a detailed and conceptual discussion of these processes see Carley and Porter 2012).

Since the composition of a scholarly topic, in terms of research fields, can be x-rayed by mapping the body of related literature by the overlay toolkit, the evolution of its field composition can also be tracked via the same model. Therefore, diffusion and integration processes underlying the dynamics of the topic may be revealed and quantified as well, by applying – and adjusting, cf. below – the measures associated with the toolkit. The basic idea is to construct a „dynamic" overlay map for the subject, that is, a series of maps picturing field composition in consecutive time periods, whereby structural changes in the history of the subject become detectable through time.

Despite the clarity of this modelling scenario, some important choices should be made, in terms of bibliometric indicators, concerning the body of literature selected for instantiating the subject under study. Depending on this choice, different aspects of both diffusion and integration processes might be captured. The three basic options are summarized in Fig. 1:

- The most straightforward method for grasping the topic dynamics is to partition related source documents into time periods (typically into years of publication), and subject each subset into analysis, separately. In terms of overlay mapping, annual changes in the field composition can be tracked within the corpus **(type B dynamics)**.

- However, diffusion (and integration) may naturally be interpreted as being exhibited through citation relations, as basic indicators of knowledge flow. This aspect is best approached by a comparative analysis of source documents and citing documents: the overlay map of each annual set is, then, compared with the overlay map for the collection of docs citing those sets. In this case, the effect („impact") of source documents on the scientific landscape may be directly observed, cohort by cohort **(type A dynamics)**.

- The third basic type of relation to explore knowledge flow patterns is to focus exclusively on the citing side. In this case, a similar time series of overlay maps might be constructed as in the first case, but instead of source documents, cohorts of citing documents are being mapped in a consecutive manner (that is, each citing cohort determined by the respective cohort of source documents being cited). This third mode of exploration addresses a yet further aspect of subject evolution, namely the dynamics of the reception of the topic within the scientific landscape **(type C dynamics).**

Based on these considerations, an innovative work has recently been set forth by Carley and Porter (2012), demonstrating the use of the overlay toolkit in mapping science dynamics. Aiming at the quantification of the degree of diffusion/integration in knowledge transfer processes, the authors introduced the index of *forward diversity* grounded in the overlay toolkit. More precisely, they used the Stirling index as a diversity measure in a novel way to characterize the structure of knowledge transfer from scholarly fields. Schematically speaking, to explore diffusion processes (1) a group of Subject Categories (SCs) was selected from the Web of Science databases as representing benchmark fields and (2) the overlay map of the record of citing documents was obtained for each benchmark SC. Upon field composition, they characterized the citing side of benchmark SCs with the Stirling index, reflecting the intellectual diversity of research relying on (referring to) the filed (SC) under study. Quantifying structural diversity of the citing side provided a simple yet powerful formalization of the extent of knowledge diffusion originating from a certain field. The kind of knowledge dynamics addressed was what we hereby call „type A dynamics": the diffusion process was followed along citation relations of source document cohorts.

**Fig. 1** *Dimensions of science dynamics typified via bibliometric relations.*

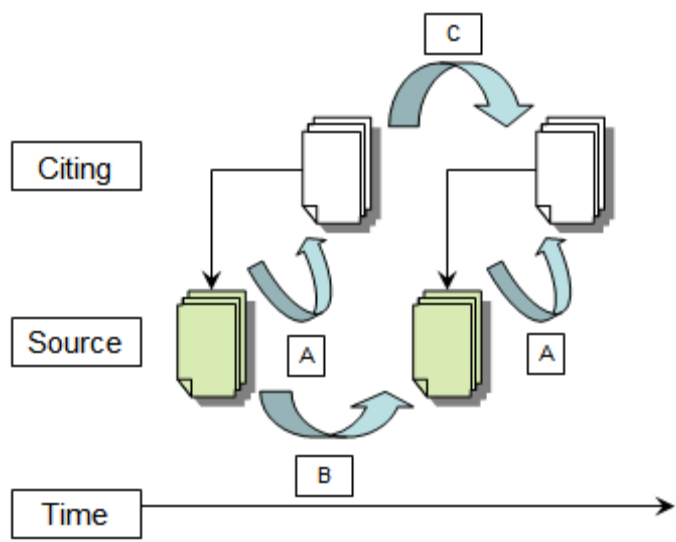

**A proposal towards dynamic diversity measurement**

The proposal formulated in the rest of this paper is closely related to the concept of *forward diversity* as a knowledge diffusion indicator, but is centering around an aspect of knowledge dynamics that has seemingly been neglected so far whenever the overlay technique was adopted. Simply put, the extent of knowledge diffusion has always been approached in an „assymetric" way. Diversity values (the Stirling index) has usually been calculated for the resulting field composition, independently of the initial one. Given e.g. „type B dynamics", obtaining diversity values for consecutive document cohorts yields a time series of diversity values (one for each timeslice), which shows how diversity changes step by step, but tells little about the underlying change of field composition. Note that the same degree of diversity may be produced by a totally different set of SCs, so that it is logically possible (though, of course, empirically rare or even unrealistic) that no (diversity) change is detected in a moment of „revolutionary transition" within the diffusion process. Similar considerations apply to the comparison of the cited vs. the citing side within a body of literature (type A dynamics): diversity on the citing side alone may inform us on the scope of knowledge reception, but to grasp *diversification* (vs. diversity) in the diffusion process, a comparison with source documents (the cited side) should also be somehow involved into the method of measurement.

In general, in order to detect genuine knowledge diffusion (either between time periods or within knowledge flow relations) the change in field composition between the source and target set seems to need some treatment. In terms of overlay mapping, the *distance* between source and target maps should be accounted for to track the shifts in the SC configuration of the topic. As the authors themselves point out, the index of *forward diversity* indicates zero diffusion when both the source and the target document sets are assigned to a single Subject Category, respectively—even when the two SCs are completely different ones. Being a definite case of field shift, one might consider it some undesired feature of the measurement, which is exactly the starting point of the present study.

To overcome this difficulty, it is tempting to turn to a tradition of measurement that is being entertained in the strongly related area of technology mapping. *Technological distance* is a concept central to the study of technology dynamics, when the basic problem is conceptually identical to the present one: to formalize the extent to which two technological profiles (of markets, of firms, or of patent portfolios etc.) differ structurally (cf. Los 2000). In technical terms, technological distance is mostly formulated as the mathematical distance of two vectors,—based on some distance metric—each representing a profile, that is, a distribution of products, patents, services over technological categories (e.g. patent classes). The main advantage of this model, as compared to diversity measurement, is that vector distances reflect an „item-by-item comparison" of profiles, therefore, the actual change in the composition of portfolios is better accounted for. On the other hand, simple portfolio vectors lack the rich variety of structural information that overlay maps (and the related diversity measurement) possess, as they do not mirror the proximity of categories themselves in the first place. Therefore, whether the difference between two distributions (i.e. technological profiles) involves redistribution among distant or close technologies cannot be told from this model alone.

**Methods**

Upon these considerations, our main aim here is to make an attempt to combine the overlay methodology, utilizing its rich measurement potential, with the concept of technological distance for measuring knowledge diffusion. Our strategy is, then, to construct an overlay-based measure of knowledge dynamics that incorporates (1) the change in the field composition between the source and the target set, respectively, and (2) the extent of potential diversification this change produces within the field composition. Our proposal to meet this challenge is what we refer to as „dynamic diversity measurement".

*Mean Overlay Distance (MOD)*

The basic idea of a dynamic diversity measure is to formulate a distance measure between two overlay maps representing the source document set and the target set, respectively. This measure is expected to take into account the similarity *between maps* and the distances *within constituent fields of the profiles* simultaneously, the latter in terms of the science map. That is, firstly, we are interested in how much the overlay map is being restructured between the source and target, and, secondly, at what distance the new field composition lies from that of the source, according to field distances indicated by the basemap.

As the previous studies, we also utilize the Stirling index to capture this two aspects of knowledge diffusion. The crucial difference is that, instead of characterizing single overlay maps, we apply the measure to the *comparison* of two maps, one of which is of the source, and the other is of the target. Given a source document set characterized with N = $n$ research fields (Subject Categories), and a target set distributed over N = $m$ Subject Categories, the proposed measure, **M**ean **O**verlay **D**istance, can be defined as follows:

$$\text{MOD} = \frac{1}{n*m} \sum_{i=1, j=1}^{n,m} p_i p_j d_{ij} \text{, whereby}$$

- $p_i$ is the relative frequency of the *i*-th Subject Category within the **source** SC-profile, $i = 1, \ldots, n$,
- $p_j$ is the relative frequency of the *j*-th Subject Category within the **target** SC-profile, $j = 1, \ldots, m$,
- $d_{ij}$ is the distance of the *i*-th (source) and the *j*-th (target) Subject Category as determined by the (common) basemap for the (both) overlays.

As can be seen from the definition, the MOD index operates on two maps the same way just as the Stirling index operates on a single map: it imposes a pairwise comparison of source and target fields (SCs), and favors those pairs, that are significant within the respective map (has high share among constituent fields), and, at the same time, cognitively distant from each other. The calculation can be conceived as the summation over the cells of a matrix of weighted source-SC-by-target-SC distances (Table 1). In other words, MOD measures both the overall (structural) difference and the (cognitive) distance between two maps. Therefore, while the previous use of the index on single overlays reports the *diversity* of SC composition, this „dynamic" extension adds the *diversification* occurred between two maps. In order to control the effect of size, the value of the „dynamic" Stirling index is normalized by the first term of the MOD formula, yielding an average of weighted distances between the two maps (hence the name *Mean Overlay Distance*).

**Table 1** *Source SC x Target SC matrix underlying the MOD index*

| Target<br>Source | $SC_{target-1}$ | (…) | (…) | $SC_m$ |
|---|---|---|---|---|
| $SC_{source-1}$ | P($SC_{source-1}$) × p($SC_{target-1}$) × d($SC_{source-1}$, $SC_{target-1}$) | (…) | (…) | (…) |
| (…) | (…) | (…) | (…) | (…) |
| $SC_n$ | (…) | (…) | (…) | p($SC_m$) × p($SC_n$) × d($SC_m$,$SC_n$) |

*Overlay Diversity Ratio (ODR)*

Given the strategy of a comparative use of the Stirling index, it is of outstanding interest how the application introduced above performs against previous uses in empirical settings. More precisely, the question is whether any process of knowledge diffusion—being modelled via the overlay methodology—shows a different picture when operationalized via the diversification-oriented MOD index versus single-map based diversity. This previous use of the Stirling measure, to simply distinguish terminologically from the MOD index, we may call **O**verlay **D**iversity (OD).

A conceptual difficulty in such a comparison lies in the very fact that the MOD index is designed for between-map usage, while the OD index applies to within-map assessment. A direct contrasting of the two measures requires a further step, whereby the *usage* of the two indices both describe the same phenomenon, namely, the transition of field composition within the evolution of a subject matter. Also, we intend to keep the original properties of the OD index for a meaningful comparison.

To meet these requirements, we introduce the concept of *Overlay Diversity Ratio (ODR)*, which is nothing more than the ratio of diversity values (ODs) for the source map and the target map, respectively. That is

$$\text{ODR} = \frac{OD_{target}}{OD_{source}}, \text{ whereby}$$

- $OD_{target}$ is the Overlay Diversity of the target set (as measured by the Stirling index),
- $OD_{source}$ is the Overlay Diversity of the source set (as measured by the Stirling index).

In verbal terms, the ODR index accounts for the relative change of the diversity in field composition between two maps. Its value equals ODR=1 in case when the transition does not effect the degree of diversity. If ODR > 1, the transition leads to an increase of diversity (a potential indication of knowledge diffusion), ODR < 1 reports a lower degree of diversity after the transition (a potential indication of knowledge integration).

*Design of the case study*

With the extended overlay toolkit discussed above, we have designed a pilot study to explore the capacity of the dynamic Stirling measure for quantifying knowledge diffusion. As our sample material, we focused on a scientific discourse with (1) a long history and (2) with a high degree of inter- and multidisciplinarity, and obtained a large-scale publication record on that topic, accordingly (cf. the *Materials* section).

Just as the proponents of *forward diversity*, we also addressed citation relations to track the potential diffusion process. In particular, (1) annual cohorts of the selected publication record were obtained, and (2) for each cohort, all papers citing its members were collected. In terms of our typology, the case study concerned **type A dynamics** (knowledge flow between the cited and the citing side). However, in order to capture the overall evolution of the topic, beyond annual sections,—that is, pairs of source cohorts and citing paper sets—we also profiled knowledge flow in a cumulative manner, by aggregating source cohorts up to each year along with the papers citing that aggregate. The rationale behind this perspective is to allow the MOD index to capture the annual extent of knowledge diffusion relative to the prehistory of the discourse at each time period, not only to the extent characteristic of a particular time period (based on the publications originating therefrom). As a consequence, with this cumulative method, both the annual values of the *Mean Overlay Distance* and the *Overlay Diversity Ratio* implicitly incorporated the measurement of **type B** and **type C dynamics** as well, inasmuch the development of field composition were captured via time-aggregation at both the cited and the citing side.

To set out formally, we have combined the above measures and methods in the following arrangements:

- *Diversification from annual sections.* For each year within the time coverage of our sample the overlay maps for annual cohorts and the related citing papers were constructed. On this basis, the *Mean Overlay Distance* between the two maps per year was obtained for monitoring the dynamics of knowledge flow in a cross-sectional perspective (that is, in each time period separately).

- *Diversification by each year (cumulative approach).* Pairs of cited–citing overlay maps were also generated by the cumulative method. In this case, the cited side for any year $Y$ was translated into an overlay map of the group of sample documents published in the year $y \leq Y$. As a consequence, each overlay map contained that of the previous years at both the cited and the citing side. It follows that annual maps (as compared to predecessors) showed the new developments (new fields) for each year in the history of the topic. The MOD index was also calculated upon this series of map pairs.

- *Diversity change by each year (cumulative approach).* In order to contrast diversification with diversity change, the *Overlay Diversity Ratio* was also applied for the cumulative (or „historical") series. In particular, the ratio of the Stirling index for the cited and the citing side maps was obtained in each time period, based on the time-aggregated maps.

**Materials: a corpus on the Species Problem**

In order to test and demonstrate the capacity of the proposed method, we applied it in an attempt to reconstruct the historical development of a rather complex discourse in biology, usually referred to as the *Species Problem*. The Species Problem can be briefly described as a historical debate on what biological species are, and as the related quest for the appropriate definition of species, or species concept for biology. With a long prehistory, dated as back as to Aristotle and Plato, including Darwin's paradigm-shifting work on the nature of species in the XIX. century (milestone #1), the debate expanded in the early XX. century, mainly due to the rediscovery of Darwin's work, and having it integrated with the early (Mendelian) genetics of the era. The new paradigm has been called the *Evolutionary Synthesis* (milestone #2). Since the Synthesis, a plethora of theories has emerged on species, resulting in a variety of competing species concepts. According to a comprehensive review of Mayden (1997), no less than 22 species concept (definitions) exhibit themselves in the contemporary literature of the subject.

Given its complexities, the Species Problem was an ideal candidate for a bibliometric analysis of —inter-, or multidisciplinary—knowledge diffusion with the proposed methodology:

(1) The roots of the discourse are centuries-old, while there are several contemporary directions of the debate (and of research) as well (cf. Hull 1988, Ereshefsky 1992).

(2) During its modern history (in the XX. century), many schools of biosystematics contributed to, and competed over the problem, involving—from a data-mining perspective—different topics: theoretical papers as well as empirical ones, the latter focusing on particular subjects of taxonomy (description of taxa). It was of outstanding interest whether the enhanced overlay toolkit was capable of identifying these knowledge transfer induced by the interaction of these schools.

(3) A nonstandard feature of the Species Problem is its complexity in terms of the contributing scholarly fields, or even disciplines. For example, a proper interaction of evolutionary systematics, on one side, and the philosophy of science (of biology), on the other side, had a significant effect on the present state of the debate. It is a good challenge for the proposed method of mapping science dynamics to capture the associated degree of knowledge diffusion being often discussed by historians of science.

To cover a representative corpus of the modern history of the discourse, bibliographic data were harvested from three databases of the Web of Science, namely, the SCI, the SSCI and the A&HCI (*Science Citation Index*, *Social Science Citation Index* and the *Arts&Humanities Citation Index*, respectively). Also in a attempt to avoid the potential exclusion of relevant works from the corpus, data retrieval was based on a topic-related query, that did not put any constraints on the set of fields, journals, authors etc. entering the sample. The query was defined to include all records related (topicwise) to any of the following terms: „species problem", „species definition", „species concept".

The resulting initial or „core" corpus included N=1605 documents for the period 1975–2012. In an attempt to gain a comprehensive historical coverage on the topic, we have iteratively extended our initial set via an in-depth analysis of aggregated references included in the core set. In the first step of the process, references from the initial corpus were processed and obtained (as source documents) from the WoS databases. This additional publication record was then added to the pool of already collected papers. We repeated this method in further iterations, until reaching a collection being fairly „closed" under the citing relation, that is, a collection that contained all the—topic-relevant—papers referred in the discourse. To assure such a convergence, references were filtered by a threshold imposed on their frequency: papers cited above this threshold were, in each round, considered *relevant* for the topic. The threshold value

10was increased (non-linearly) for each iteration, based on the assumption that the farther we get, along a series of references, from the core set of papers (in terms of corpus generations), the less related references will be to the topic. Interestingly, with this setting, the procedure converged in the third generation of papers, indicating that almost all relevant references were present after two iterations. Finally, for the discourse of the species problem, we arrived at a final record of approximately 5700 papers (the main statistics of the procedure are summarized in Table 2.)

**Table 2.** *Statistics of iterative corpus collection on the Species Problem based on WoS databases*

| Iteration | No. of source documents | No. of references | No. of unique references | Threshold value | No. of relevant references (retrievable) |
|---|---|---|---|---|---|
| Initial corpus | 1605 | 93 943 | 50 668 | 3 | 3223 |
| 2. generation | 3223 | 155 742 | 62 574 | 10 | 851 |
| 3. generation | 851 | 14 991 | 5305 | 10 | 2 |
| **Total** | **5679** | | | | |

To prepare the monitoring of knowledge diffusion along this large-scale longitudinal bibliography, we have organized the final corpus into a citation network. The large directed graph obtained from the document set consisted of all papers included in the full corpus as its nodes; edges represented the (direct) citing relation between any two documents. The rest of the paper reports the comparative (longitudinal) analysis of this network via our extended overlay toolkit.

**Results and discussion**

Results of applying the extended overlay toolkit to the citation network of the species problem are presented in Fig 2–4. below. The three plots display the outcome of the three experimental designs discussed above, respectively. Fig. 2 shows the time series of *Mean Overlay Distance* values over the time period 1975–2011, measured between (separate) annual source cohorts and their citing environment within the corpus (diversification induced by annual sections of the species topic). Our main focus is Fig. 3, whereby the MOD index is calculated via the cumulative method, between the cited side and the citing side being aggregated up to each year within the timespan (diversification *within* the species *topic* by each year). Finally, the longitudinal measurement of the *Overlay Diversity Ratio* is set out in Fig. 4, for the purposes of comparison with the MOD measure. The underlying citation relation was also obtained by the cumulative method, that is, the *Diversity Ratio* was established between cohorts published in year $y \leq Y$ and the citing nodes for $y$ = 1975, …, 2011.

The most striking outcome of this three-way comparison is, in general, that the three curves report a (pairwise) different tendency on the knowledge sharing process, or, more precisely, show a different aspect of the very same process. The cross-sectional perspective (Fig. 2) exhibits an initial fluctuation in the MOD index (up to the late eighties), becoming much more moderate, almost steady in later periods, with a sudden peak around 2010. Overall (as a potential smoothing regression of the empirical values would reveal), a descending tendency can be observed in this dimension of citation-based knowledge transfer. This descending tendency is much more heavily present (and differently shaped) in the cumulative perspective (Fig. 3). It can be viewed as evidencing, basically, the effect of novel fields entering the topic each year (being absent in previous years) at both the citing and the cited side. MOD values in this application follow a power law-like curve: diversification (through citations) is quickly decreasing up to the early eighties, which tendency continues but slows down in the 80's, and a very low, slightly falling but basically steady level is observable in the last two decades. On the other hand, a radically different picture emerges from measuring the change of diversity between time periods (Fig. 4). The ODR index, apart from an initial small oscillation, quickly raises above ODR = 1 and remains within the interval between 1–1.2. In other words, as opposed to the MOD index, knowledge transfer results in an increased diversity of research fields throughout the whole timespan of topic development.

As a brief interpretation of these results, the species problem may be characterized as having a vivid or „revolutionary" period in the 70's–80's opening up interfaces between various and relatively distant fields of research, as evidenced by the MOD index. The cross-sectional analysis suggests that different fields entered the scene in subsequent years, with far-reaching impact relative to each year. The cumulative approach adds, however, that field composition quickly became „saturated", that is, by accumulating fields along the timeline, the cited and the cited side took an increasingly similar structure. This interpretation is in accord with historiography: according to reconstructions on the history of the topic, a fundamental drive behind the modern debate on the species concept was a thesis from the philosophy of science originating from the mid-seventies and disputed mainly throughout the eighties, that became widely accepted and assimilated within theoretical biology. This so-called *individuality thesis*—stating that species are ontological and methodological individuals—therefore, invited fields such as the *history and philosophy of science* into the discourse otherwise dominated by the life sciences (cited side), and, being a rather influential one, propagated through a variety of fields (w.r.t. the citing side). This extended scope, once being emerged, remained characteristic of the topic in later periods, resulting both (1) little sign of „additional" diffusion and (2) high degree of diversity inherited from this early „boom" of subject areas, as clearly reflected by overlay diversity ratios (ODRs).

The comparison between diversity ratios (Fig. 4) and—cumulative—overlay distances (Fig. 3) is especially intriguing since both series have been obtained from the same set of cumulative maps. In order to gain a deeper insight into the very process behind the tendencies captured via our proxies, we visualized knowledge transfer at two selected time periods. Respective overlay maps of the cited and the citing side are presented in Fig 5 for the year 1976 and 2001. These two timeslices are quite illustrative as witnessing rather different degrees of diversification (MOD index), but highly similar values of diversity change (ODR index). By consulting the related maps, however, an explanation presents itself.

12Source documents published in 1976 were distributed in Subject Categories from mainly the life sciences—Biomedical Sciences, Ecological Sciences, Agricultural Science, Infectious Diseases as disciplines—accompanied with some „non-life" hard sciences (e.g. Geosciences). An area positioned farther from these fields were „Social Studies", increasing the distance-based Stirling index for the profile. The associated map indicates that citing papers span a similar field composition, but further Subject Categories, both in the same areas and also in farther regions widen the spectrum of reception: most importantly, a set of fields „mediating" between the „social sciences" and the „natural sciences", namely, Cognitive Sciences enter the scene with two SC in the middle of the map (increasing the effect of distance in the measurement). In the „social pole", Business and Management Sciences also pop up. Turning to 2001, both the source map and the target map are much more diversified in this late period, but also much more similar to each other: though novel and relatively distant SCs show up in the citing environment from Computer Science and the collection of „Economics, Politics and Geography", their share is almost ignorable (below 0,01 percent), so their contribution to the overall share-weighted distance from the source map is almost invisible.

To sum up these effects, knowledge transfer in both years leads to a higher diversity of fields (in terms of the Stirling measure), keeping the ratio of diversities above 1. However, despite of this increment, the overall distance of the citing composition is considerably higher in the early period (described by fewer SCs) than in the later year under study (whereby SCs are abundant). Hence the parallel fall in the MOD measure. This result, beyond explaining the values presented within the time series, provides justification for the use of both approach, as capturing different aspects of the diffusion process.

**Fig. 2** *Development of the MOD index comparing annual sections and their citing environment*

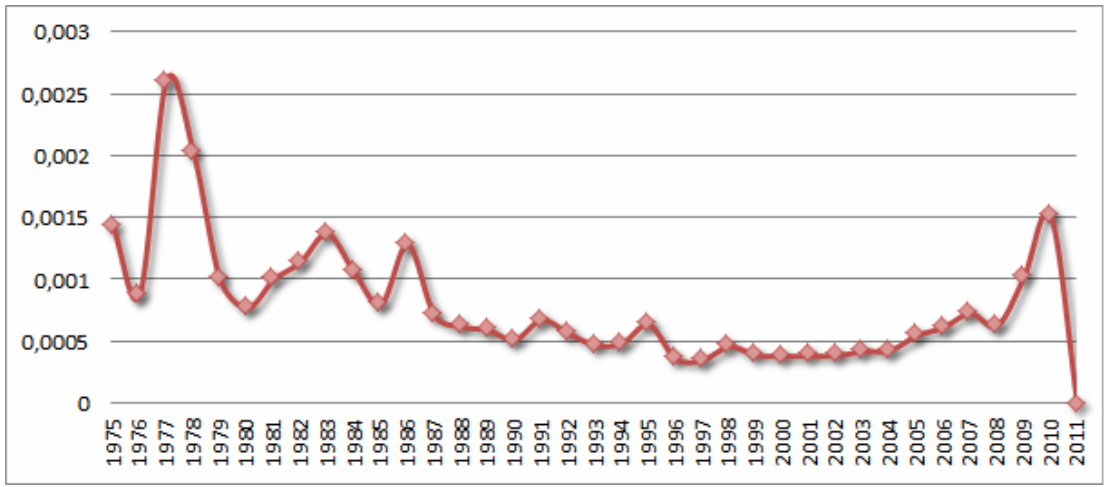

**Fig. 3** *Development of the MOD index comparing accumulated papers wit their citing environment*

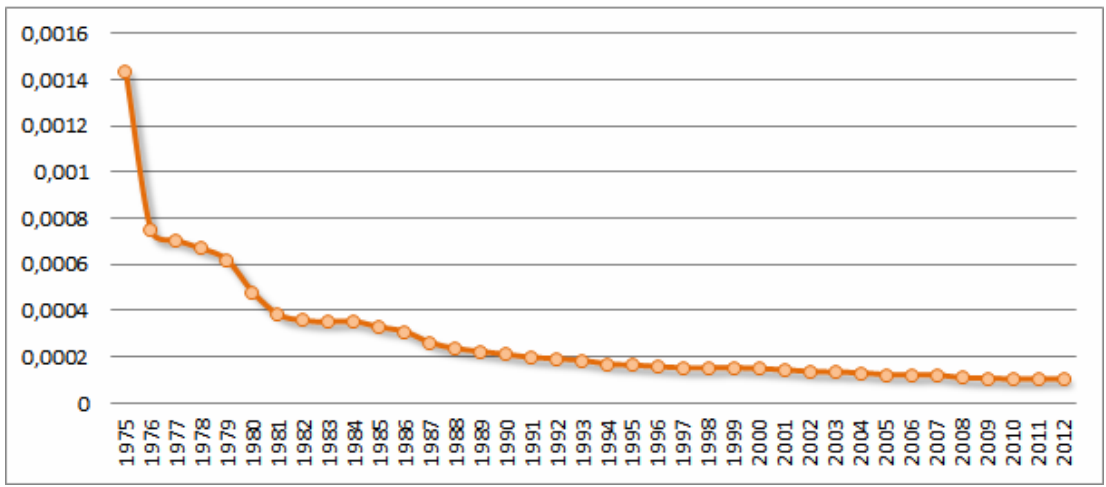

**Fig. 4.** *Development of the ODR index comparing the diversity of accumulated papers up to each year with the diversity of their citing environment*

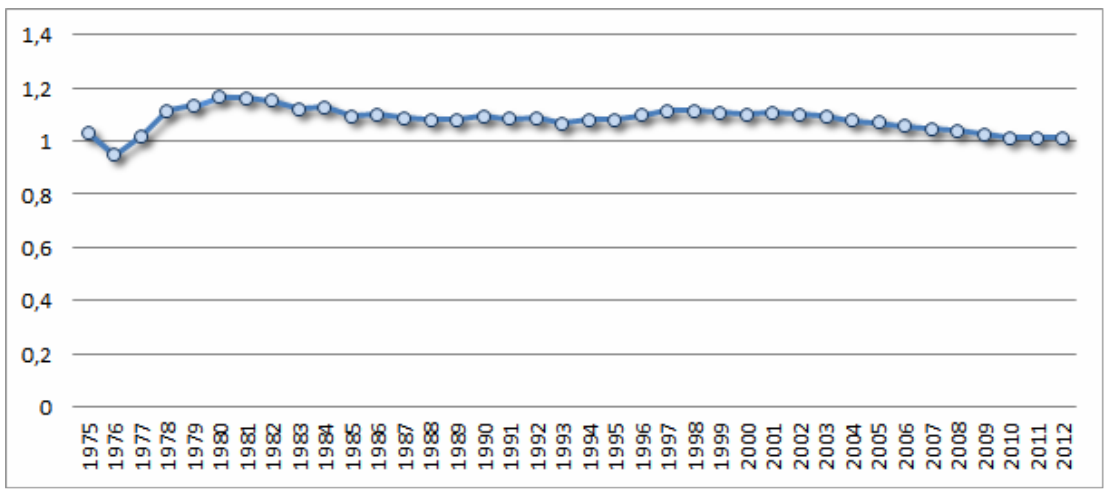

**Fig 5** *The overlay maps for two selected years visualizing knowledge diffusion between the cited (left) and citing side (right): 1976 (top), 2001 (bottom)*

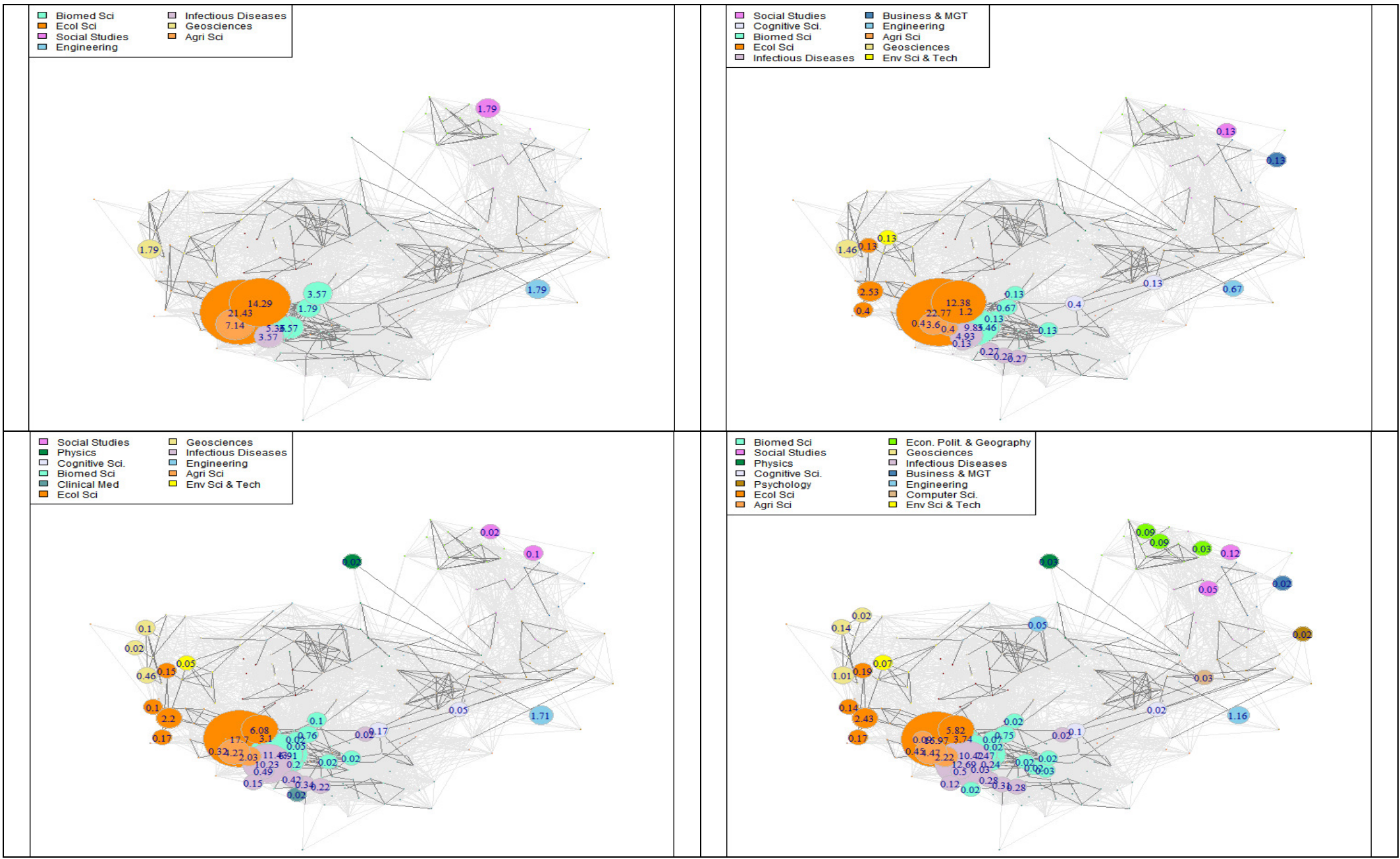

**Conclusion**

The work reported in this paper was aimed at investigating novel uses of the so-called *science overlay toolkit* for mapping science dynamics. In particular, focusing on the process of knowledge diffusion, we have modelled the development of a scientific topic via dynamic overlay maps, upon which two new variants/derivatives of the popular Stirling Index were introduced. The *Mean Overlay Distance* (MOD) index has been argued to account for those ingredients of diversification/integration—in terms of field composition—that previous uses of the Stirling index have failed to capture so far, as was shown by comparing the performance of the MOD index with recent approaches. As a case study, we demonstrated our approach on a large-scale corpus covering a historical topic from the life sciences, namely, the *species problem.*

An important conclusion to be drawn from this comparative study is that, when it comes to mapping the patterns of knowledge transfer, the set of investigated measures exhibit a complementary relation, addressing different dimensions of science dynamics each. To the more, while the MOD measure seems to better reveal the amount of „shift" in field composition (size), the ODR measure (reflecting recent applications of the Stirling index as a diversity metric) indicates the direction of this transition (whether it is a case of diversification or integration). The MOD and the ODR measure, therefore, jointly define the „full vector" of the change along the timeline.

An equally relevant feature of the toolkit proposed in this study is its capability to model a variety of dimensions of science dynamics, accompanied by its applicability in a wide range of research domains. As a direct follow-up of the present use case we are currently working on the modelling of the development of individual, institutional etc. research profiles referred to as „thematic mobility" in the context of monitoring academic careers, which is a case of type B dynamics x-rayed with the extended toolkit. A still further case is an evaluative application of the new indices in assessing citation impact not only through quantities (times cited), but through „type A dynamics", that is, by quantifying also the scope of citation impact over the scientific landscape.

# Acknowledgement

This paper was supported by the János Bolyai Research Scholarship of the Hungarian Academy of Sciences; the European Union and the European Social Fund through project [FuturICT.hu](FuturICT.hu) (grant no.: TÁMOP-4.2.2.C-11/1/KONV-2012-0013); and the European Commission under the FP7 Science in Society Grant No. 266588 (SISOB project).